\documentclass[aps,prl,twocolumn]{revtex4}
\usepackage{epsfig}
\usepackage{amsmath}
\usepackage{ulem}

\begin{document}
\title{Detection of weak emergent broken-symmetries of the Kagom\'e antiferromagnet from Raman spectroscopy}
\author{O.~C\'epas,$^1$ J.~O.~Haerter,$^{1,2}$  C.~Lhuillier$^1$}
\affiliation{
$1.$  Laboratoire de physique th\'eorique de la mati\`ere condens\'ee, UMR 7600 C.N.R.S., Universit\'e Pierre-et-Marie-Curie, Paris VI, France. \\
$2.$  Department of Physics, University of California, Santa Cruz, CA 95064, USA.
}
\date{\today}

\begin{abstract}
  We show that the magnetic Raman response of a spin-liquid is
  independent of the polarizations of the light for triangular
  symmetries. In contrast, a ground-state that has a broken symmetry
  shows characteristic oscillations when the polarizations are
  rotated. This would allow to detect weak broken symmetries and
  emergent order-parameters. We focus on the Kagom\'e antiferromagnet
  where no conventional long-range order has been found so far, and
  present the Raman cross-section of a spin-liquid and a valence bond
  crystal (VBC) using a random phase approximation.
\end{abstract}

\pacs{PACS numbers:} \maketitle

Interacting spins on the Kagom\'e lattice are particularly interesting as a
possible realization of a spin-liquid ground state. The recent discovery of a
magnetic oxide, ZnCu$_3$(OH)$_6$Cl$_3$ \cite{Shores}, with the geometry of the
Kagom\'e lattice has triggered renewed interest in this field
\cite{Today}. All recent experiments in this compound point to the absence of
long-range order, at least of conventional form, and at temperatures well
below the superexchange coupling: a persistent dynamics at the lowest
temperatures \cite{Mendels,Ofer}, no Bragg peaks in neutron scattering
\cite{Helton}. This is all consistent with the prediction of a non-magnetic
ground state by exact diagonalizations of the Heisenberg model \cite{Leung,
  Lecheminant,Lhuillier}. Although other interactions (such as that of
Dzyaloshinski-Moriya symmetry) may play a role in this compound \cite{Rigol},
the nature of the ground-state of the Heisenberg model is currently under
debate. It could be a spin-liquid which does not break any symmetry
\cite{Ran}, or a valence-bond-crystal (VBC) that breaks translation symmetry
\cite{Marston,Zeng,Maleyev,Budnik,Senthil,Huse,Sachdev}.  In the latter cases,
however, the unit-cells turn out to be rather large and the modulation of the
correlations must be weak, otherwise a large spin-gap would exist.  This makes
the use of exact spectra to identify the broken symmetry a difficult task
\cite{Misguich}, and no definite proof has emerged so far on the theoretical
side. The question then arises as to whether it could be possible to detect
such a weak-broken symmetry experimentally. Detecting the resulting
emergent order-parameters is indeed not easy, for lack of direct experimental
probes: indirect ways were suggested for the chiral order-parameter, for
instance, through polarized neutron experiments \cite{Mal} but those do not
probe the singlet sector of VBC's.

In this Letter, we argue that the Raman scattering of light by spin
excitations provides a way to distinguish between a spin-liquid and a
broken-symmetry state. In particular, it allows one to extract emergent
order-parameters in systems with no conventional magnetic long-range order.
Indeed, we have found that the Raman cross-section is
\textit{polarization-independent} for a spin-liquid governed by a Heisenberg
Hamiltonian in triangular geometries, whereas it acquires a characteristic
\textit{polarization dependence} for a broken-symmetry state, with an
amplitude proportional to the emergent order-parameter.

Magnetic Raman scattering is a two-photon process: the (polarized)
electric-field of light
forces the exchange of two electrons (and hence their spins), thus creating a
spin excitation in the system. The cross-section of this
process is given (at zero temperature) by  \cite{Loudon, Shastry,noteRaman}
\begin{eqnarray}
S(\omega,\hat{\mathbf{e}}_{in},\hat{\mathbf{e}}_{out}) = \sum_f |\langle
f|H_R|0 \rangle|^2 \delta(\omega -\omega_f) \label{cs} \\ H_R  =
\sum_{<i,j>} (\hat{\mathbf{e}}_{in} \cdot \mathbf{r}_{ij})
(\hat{\mathbf{e}}_{out} \cdot \mathbf{r}_{ij}) \mathbf{S}_{i} \cdot
\mathbf{S}_j
\label{Raman}
\end{eqnarray}
(\ref{cs}) depends explicitly on the polarizations of the incoming and outgoing photons,
$\hat{\mathbf{e}}_{in,out}$ through the dipolar factors  $\hat{\mathbf{e}}
\cdot \mathbf{r}_{ij}$, where $\mathbf{r}_{ij}$ is the bond
vector of nearest-neighbor sites. The actual polarization dependence results
from the symmetries: if the ground-state $|0 \rangle$ is a spin-liquid (that does not break any
lattice symmetries), one can predict the polarization-dependent terms according
to the symmetries of the excited states $|f \rangle $ (of energy
$\omega_f$). For instance, in square geometries, all scattering channels
depends on polarizations \cite{Devereaux}. In triangular geometries, however,
the cross-section turns out to be polarization-independent for a spin-liquid,
as is shown below using symmetry arguments. Any departure from isotropy
therefore signals a broken-symmetry state.

The problem of the energy dependence of the Raman cross-section (\ref{cs})  is 
also of particular interest: what form of singlet response do we expect for a non-magnetic ground-state? In a
conventional ordered-state with $\langle \mathbf{S}_i
\rangle \neq 0$, the response (\ref{cs}) is dominated  by 
two-magnon excitations at low energy and can be calculated accordingly \cite{Devereaux}. Since this cannot obviously  apply to a
state with $\langle \mathbf{S}_i \rangle = 0$, we have developed the simplest
random phase approximation (RPA) for the singlet dynamics of putative
spin-liquid or VBC states.  RPA is a rather drastic approximation that cannot
reproduce the large singlet sector of the Kagome lattice
\cite{Lecheminant,Lhuillier}. Nonetheless it gives simple well-defined
excitations that illustrate the polarization properties explicited above and which
 can be further tested for various lattices.

 First, we discuss general selection rules and the polarization-dependent terms
of the
cross-section using symmetry arguments.
For this, the Raman operator (\ref{Raman}) is decomposed in irreducible
tensors. The point group of the Kagom\'e lattice at
$\mathbf{k}=0$ is that of the triangular lattice, $C_{3v}$ (there is an additional parity since each site is an inversion center, but it does not play any role in the following).  $C_{3v}$
has two one-dimensional irreducible representations (IR) $A_1$ and
$A_2$ and one two-dimensional IR $E$. The Raman operator does not have
a projection onto $A_2$ so we have
\begin{equation}
H_R = (\hat{\mathbf{e}}_{in} \cdot \hat{\mathbf{e}}_{out}) O_{A_1} +  \mathbf{M} \cdot \mathbf{O}_E
\end{equation}
where $O_{A_1}$ and $\mathbf{O}_E$ are irreducible tensors that
transform according to $A_1$ and $E$ respectively. $O_{A_1}$ is in
fact nothing but the Heisenberg Hamiltonian so there is no
scattering by $A_1$ excited states.  This gives a first selection
rule: if the ground state is a spin-liquid (with $A_1$
symmetry, the most general case), only the excited
states belonging to $E$ (they are twice degenerate) are
Raman-active. The vector $\mathbf{M}$
contains the polarization properties and is expressed by $
\mathbf{M} = \sum_{i} (\hat{\mathbf{e}}_{in} \cdot \hat{u}_i)
(\hat{\mathbf{e}}_{out} \cdot \hat{u}_i) \hat{u}_i $, where the sum
runs over the bonds of the unit-cell, and $\hat{u}_i$ are the
unit-vectors along these bonds $\hat{u}_1=(1,0)$,
$\hat{u}_2=(-1/2,\sqrt{3}/2)$, $\hat{u}_3=(-1/2,-\sqrt{3}/2)$.
$\mathbf{M}$ can be simply reexpressed with its coordinates
$\mathbf{M} \propto (\cos (\theta_{in}+\theta_{out}), \sin
(\theta_{in}+\theta_{out}))$, where $\theta_{in}$ and $\theta_{out}$
are the angles of the polarization vectors of the incoming and
outgoing photons with respect to the $x$-axis. The Raman
cross-section (\ref{cs}) at zero temperature is reduced to the
Fourier transform of $\langle 0| H_R(t)H_R(0) | 0 \rangle =
\sum_{\alpha \beta} M^{\alpha} M^{\beta} \langle 0| O_E^{\alpha}(t)
O_E^{\beta}| 0 \rangle$. We use the irreducible decomposition of
$\langle0| O_E^{\alpha}(t) O_E^{\beta}|0\rangle$ and that of
\begin{equation}
M^{\alpha} M^{\beta} = \frac{1}{2} \mathbf{M}^2 \delta_{\alpha \beta}+  \frac{1}{2} \sigma^z_{\alpha \beta} (M^{x 2}-M^{y 2}) +  \sigma^x_{\alpha \beta} M^xM^y
\label{M2}
\end{equation}
where the $\sigma^{x,z}$ are the Pauli matrices.  The first term
belongs to $A_1$ and the last two terms to $E$, there is no
projection onto $A_2$. Each term of (\ref{M2}) is multiplied by the
matrix element $ \langle 0| O_E^{\alpha}(t) O_E^{\beta}| 0 \rangle $
of the same symmetry. This matrix element is zero or not depending
on the symmetry of the ground state.  \textit{(i)} The ground-state
is assumed to be a spin-liquid, \textit{i.e.} it does not break any
crystal symmetry (the wave-function $|0\rangle$ transforms according
to, \textit{e.g.},  $A_1$). In this case the Raman cross-section
reduces to the $A_1$ terms in the decomposition,
$\mathbf{M}^2 \langle 0| \mathbf{O}_E(t) \cdot \mathbf{O}_E| 0
\rangle$. Remarkably, $\mathbf{M}^2$ does not depend on
$\hat{\mathbf{e}}_{in}$ or $\hat{\mathbf{e}}_{out}$. Therefore, the
Raman cross-section of a spin-liquid is rotationally invariant:
\begin{equation}
\mbox{\textit{Spin liquid : }} \hspace{0.5cm} S(\omega,\hat{\mathbf{e}}_{in},\hat{\mathbf{e}}_{out})= A(\omega) \nonumber
\end{equation}
 This is a special property
of the triangular symmetry that cannot be general:
 for example, on the square lattice all scattering channels are
 polarization-dependent \cite{noteSquare}.
 \textit{(ii)} Now suppose that the ground state
 spontaneously breaks a symmetry of the crystal. For instance we could
 have a N\'eel state or a VBC. In both cases the wave-function contains a
 superposition of degenerate states belonging to different IR, $| 0 \rangle= a |A_1\rangle + b |A_2\rangle + c |E\rangle$. As a consequence,
 the Raman cross-section contains the additional terms of (\ref{M2}),
 $M^{x 2}-M^{y 2}=\cos [2(\theta_{in}+\theta_{out})]$ and $M^xM^y=\sin
 [2(\theta_{in}+\theta_{out})]$ with a prefactor that depends on the cross-terms $a c^*$, etc.:
\begin{eqnarray}
&& \mbox{\textit{Broken lattice symmetry : }} \hspace{0.5cm} \nonumber \\ && S(\omega,\hat{\mathbf{e}}_{in},\hat{\mathbf{e}}_{out})= A(\omega) + E(\omega) \cos [2(\theta_{in}+\theta_{out})+\phi_{\omega}] \nonumber
\end{eqnarray}
The result depends explicitly on $\theta_{in}$ and $\theta_{out}$
 through the cos term ($\phi_{\omega}$ is a phase factor independent
 of $\theta_{in}$ and $\theta_{out}$). The amplitude of the
 oscillation $E(\omega)$ is related to the order-parameter of the
 broken-symmetry ground-state and it is weak if the symmetry is weakly
 broken. It appears as a measure of the cross-terms of the ground
 state wave-function.  Furthermore if the symmetry is broken, we can
 possibly see individual excited states $| f \rangle $ with a transition
 probability given by $|\langle f| \mathbf{M} \cdot \mathbf{O}_E |0
 \rangle|^2 \sim \cos^2(\theta_{in}+\theta_{out}+\phi_f)$. If it is
 possible to resolve individual peaks experimentally then the
 variation with polarizations is strong, otherwise the sum of them
 reduces to the expression above.

  In conclusion, Raman spectroscopy appears as an appropriate probe to
 show whether the ground-state breaks the crystal symmetries or not:
 if the response is rotationally invariant we can conclude that the
 ground state is a spin-liquid \cite{note1}; if not the ground-state
 should break the symmetry of the crystal. In this case, the amplitude
 of the modulation gives access to the order-parameter. In the case of
 the Kagom\'e system where no ordered moment has been found, this may
 help experimentally to distinguish between a spin-liquid and a VBC
 for instance.

This discussion also applies to the triangular lattice.  Incidentally,
 the Raman response of the Heisenberg antiferromagnet on the
 triangular lattice has been calculated recently using exact
 diagonalizations of small clusters \cite{DevereauxT}. It is important
 to stress that this result of \cite{DevereauxT} cannot be seen as reflecting the Raman
 response of the ordered state in the thermodynamic limit: the
 oscillation has been missed because in a finite-size system the
 ground state belongs to the trivial representation of the group.  One
 would need to construct the semi-classical N\'eel state by summing
 wave-functions of different IR with the correct amplitudes.  The
 selection rules and the polarization dependence should then reflect
 the properties given above.

In order to have more precise predictions for the Raman spectrum, we now
present the results of a random phase approximation.  The approach consists of
writing down hierarchical equations of motion for the singlet operator
$\mathbf{S}_{i}(t) \cdot \mathbf{S}_{j}(t) $ (with $i$ and $j$ anywhere on the
lattice) and the four-spin susceptibility associated with (\ref{cs}). The
closure of the hierarchy is done in such a way as to use
two-point correlation functions, $\langle \mathbf{S}_i \cdot \mathbf{S}_j
\rangle$ as decoupling parameters instead of the local magnetizations $\langle
\mathbf{S}_i \rangle$ that are assumed to vanish in the system. This involves
writing down the equation of motion to second-order in the
time-derivative. The closed equation of motion then contains c-numbers that
are the Fourier components of $\langle \mathbf{S}_{i} \cdot \mathbf{S}_{j}
\rangle$. These numbers are unknown and must be determined self-consistently
using sum-rules.  This is similar to the Kondo-Yamaji decoupling in
one-dimensional systems \cite{KY}, and can be viewed as a random phase
approximation. For the Kagom\'e lattice with no broken symmetry, the
self-consistent parameters were previously determined
\cite{Lacroix,footnote}. In addition, we can construct a phenomenological
theory of a VBC that breaks the spatial symmetries, simply by imposing an
\textit{ad hoc} modulation of the $\langle \mathbf{S}_{i} \cdot \mathbf{S}_{j}
\rangle$ over the previous parameters. For both systems, the equations of
motion are solved numerically in the reciprocal space and the Raman
cross-section (\ref{cs}) is extracted at $\mathbf{k}=0$ (the system contains
up to 2700 sites).
\begin{figure}[htbp]
\centerline{
\psfig{file=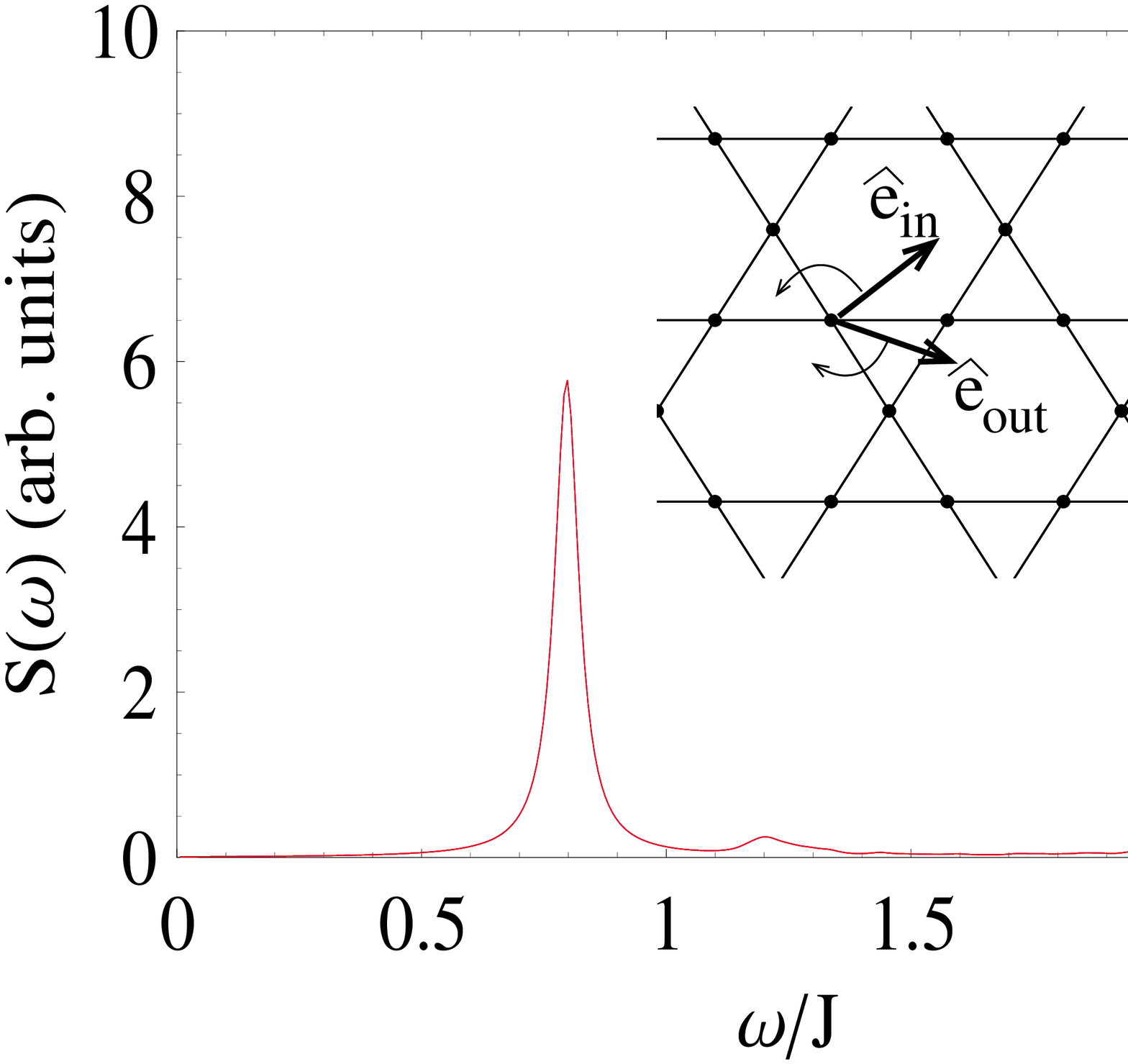,width=8.5cm,angle=0}
}
\caption{(color online). Magnetic Raman cross-section of the Kagom\'e lattice calculated within RPA, assuming a spin-liquid ground-state. The response does not depend on the orientation of the incoming and outgoing photon polarizations, $\hat{\mathbf{e}}_{in}$ or $\hat{\mathbf{e}}_{out}$. Each individual delta function is slightly broadened with a Lorentzian.}
\label{Raman1}
\end{figure}

 In Fig.~\ref{Raman1}, we give the result of the RPA calculation of
$S(\omega,\hat{\mathbf{e}}_{in},\hat{\mathbf{e}}_{out})$ for the Kagom\'e lattice, assuming that $ \langle
\mathbf{S}_{i} \cdot \mathbf{S}_{j} \rangle$ do not break the
crystal symmetries (we replace the $\delta$ functions in Eq.~(\ref{cs}) by Lorentzians). We find that the
spectrum consists of several peaks, the intensity of which decreases
strongly when the energy increases. The main intensity is in fact
quite concentrated in a single mode at about $0.8J$. The result is
fully in agreement with the group theory arguments given above: (i)
all Raman-active modes are twice degenerate, so their wave-functions
can be labelled with the IR $E$, (ii) we have rotated the photon
polarizations $\hat{\mathbf{e}}_{in}$ and $\hat{\mathbf{e}}_{out}$ and
have found that the intensities in Fig.~\ref{Raman1} do not change at
all. By comparing with preliminary exact results of the dimer
dynamical response of small clusters \cite{Lauchli}, we believe that
the response in the single peak at $\omega \sim 0.8J$ can in fact be
broadened. This would be interesting as an indication that the present
simple excitation is coupled to other low-energy singlet modes and
would decay accordingly. Of course, the present approach has neglected
this effect in closing the hierarchy of higher-order Green's
functions.

\begin{figure}[htbp]
\centerline{ \psfig{file=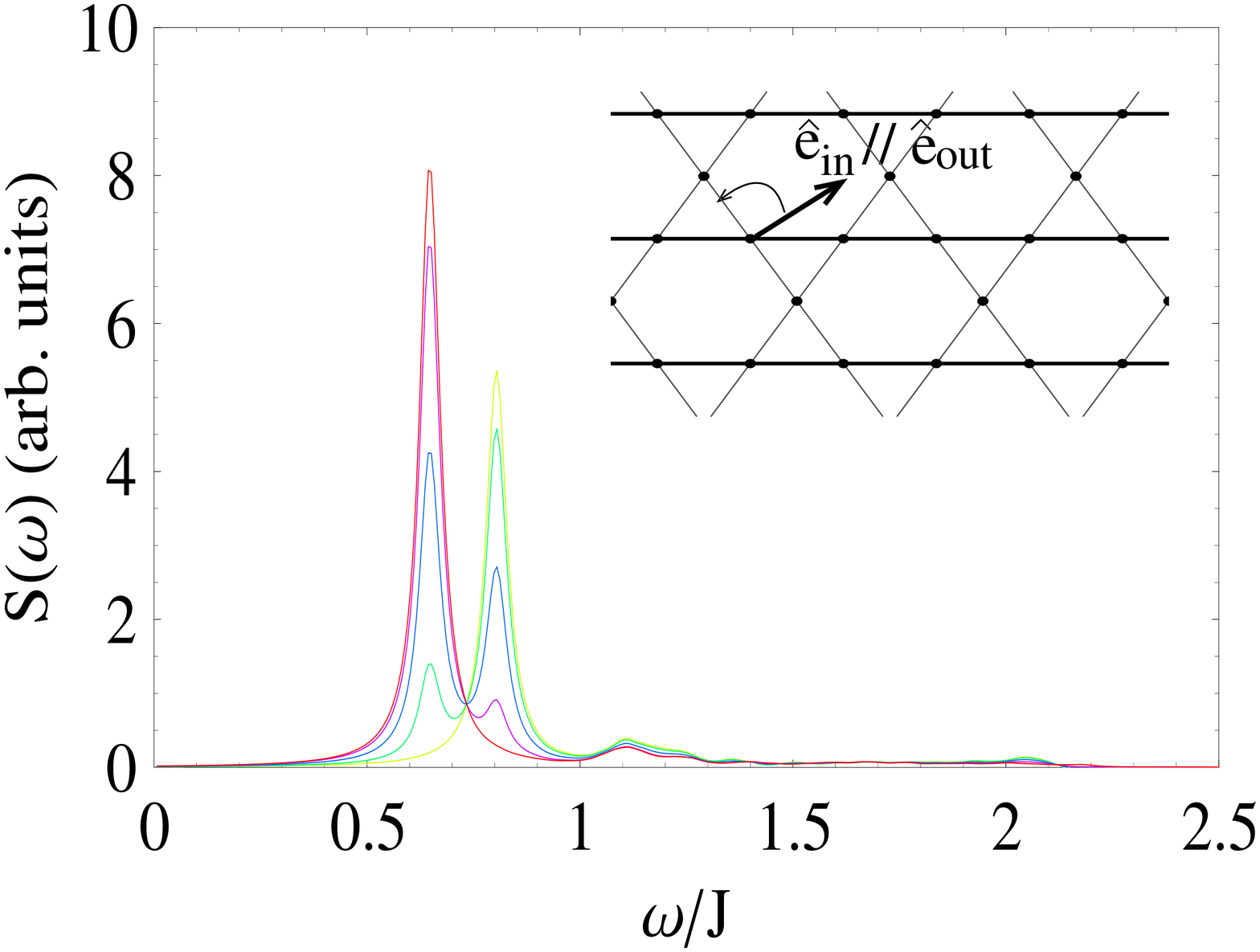,width=9cm,angle=0}}
\caption{(color online). Magnetic Raman cross-section of the Kagom\'e
lattice with weak broken-symmetry, within RPA. The ground-state is
supposed to be the valence bond crystal depicted in the inset,
that breaks the rotation symmetry (the correlations are assumed to be stronger on horizontal bonds). Different curves correspond to
different orientations of the polarization vector $\hat{\mathbf{e}}_{in}
(\parallel \hat{\mathbf{e}}_{out}$).}
\label{Raman2}
\end{figure}

Furthermore, we now choose a VBC ground-state with a spontaneous
broken symmetry in the two-point correlation functions. For simple
illustration, we impose the same weak perturbation of slightly
stronger $\langle \mathbf{S}_{i} \cdot \mathbf{S}_{j} \rangle $ on all
horizontal bonds (30 $\%$ stronger in the following calculation) (see
inset of Fig.~\ref{Raman2}).  Since the equation of motion within RPA
breaks the $C_{3v}$ symmetry, the two-fold degeneracy of the
$E$-states is lifted and we have pairs of peaks. The intensities now
display strong orientation dependence when $\mathbf{e}_{in} \parallel
\mathbf{e}_{out}$ is rotated in the plane, as shown by the different
curves in Fig.~\ref{Raman2}. The intensities of the two peaks change
mainly like $\cos^2 2 \theta$ ($\theta_{in}=\theta_{out}=\theta$) for
one component and $\sin^2 2 \theta$ for the other with different
prefactors as long as the order-parameter is non-zero.  If the peaks
cannot be resolved experimentally, the cross-section measures the sum
of the two intensities that therefore reduces to $A(\omega_0) +
E(\omega_0) \cos 4 \theta $, in agreement with the general argument
given above. It should be emphasized that even if the symmetry is
weakly broken, the sum still displays a weak characteristic
oscillation, the amplitude of which gives access to the
order-parameter of the broken symmetry state \cite{noteconclusion}. We
have shown this on the simplest example of a VBC, but we believe this
will remain true for more complicated superstructures with $\mathbf{k}
\neq 0$, such as those suggested in the literature
\cite{Zeng,Marston,Maleyev,Budnik,Senthil,Huse,Sachdev}. It is
interesting to note that a VBC will be presumably accompanied by
lattice distortions of the same symmetry, which can be tested independently by X-rays.

We conclude with a simple selection rule for Raman spectroscopy in
triangular geometries: if the ground state is a spin-liquid, the Raman
response is independent of the polarizations of the incoming and
outgoing photons. However if the ground state has a broken lattice
symmetry, it should depend, in most cases \cite{note1}, on the
polarizations of the light.  The dependence is given by $\cos
[2(\theta_{in}+\theta_{out})+\phi]$ with an amplitude that measures
the strength of the emergent order-parameter. Raman spectroscopy can
therefore test directly the presence of such broken symmetries.  This
may help in clarifying the ground state of the Kagom\'e system
ZnCu$_3$(OH)$_6$Cl$_3$ and especially to discriminate between a real
spin-liquid and a valence-bond-crystal.  Other particularly
interesting candidates are the possible spin-liquids on triangular
lattices: the organic material $\kappa$-(BEDT-TTF)$_2$Cu$_2$(CN)$_3$
(it is nearly an isotropic triangular lattice) \cite{Kanoda}, or
NiGa$_2$S$_4$ \cite{Nakatsuji}.  Here again the issue of emergent
broken-symmetry is of particular interest.

We thank A.~Laeuchli and P.~Lemmens for giving us accounts of their results
prior to publication, and Y.~Gallais and A.~Sacuto for discussions.
O.~C. would like to thank also G.~Bouzerar for pointing out Ref.~\cite{KY} and
for discussions, and the ILL theory group. J.~H. was supported by UCSC and IUF
grants for this project.


\begin{thebibliography}{99}
\bibitem{noteRaman} We will neglect corrections
 to (\ref{Raman}) coming from small spin-orbit
coupling \cite{Loudon}, or smaller (at least away from resonance) higher-order processes \cite{Shastry}.
\bibitem{noteSquare} For the square lattice, in the A$_{1g}$, A$_{2g}$,
  B$_{1g}$, B$_{2g}$ scattering channels, for instance, the
  polarization dependence would be given respectively by $\cos^2
  (\theta_{in}-\theta_{out})$, $\sin^2
  (\theta_{in}-\theta_{out})$,  $\cos^2
  (\theta_{in}+\theta_{out})$, $\sin^2
  (\theta_{in}+\theta_{out})$ \cite{Devereaux}.
\bibitem{note1} A word of cautious : if the ground-state is a
superposition of $A_1$ and $A_2$ only (and not $E$), the symmetry is broken,
 but since (\ref{Raman}) does
not have  projections on $A_2$, the cross-section will not depend on
$\hat{\mathbf{e}}_{in,out}$ either. This may be the case in
the VBC of refs.~\cite{Senthil,Maleyev}, but 
certainly not the case of the N\'eel state nor of the VBCs of
refs.~\cite{Budnik,Huse} which remain the most plausible VBC
candidates to describe the Kagom\'e antiferromagnet.
\bibitem{footnote} We have used the parameters \cite{Lacroix} $\langle
\mathbf{S}_{i} \cdot \mathbf{S}_{j} \rangle$ = -0.213 for the first
shell (nearest neighbors), 0.042 for the second, 0.037 for the third
and zero for the next ones. This is a good approximation of the exact values obtained on small clusters \cite{Leung}. We have added the vertex
parameter $\alpha_v=1.42$ \cite{Lacroix}, that ensures the sum rules
to be fulfilled but had to decrease its value by $15 \%$ to avoid
complex frequencies (in principle these parameters must be
recalculated using the new excitations).
\bibitem{noteconclusion} For the simplest VBC of Fig.~\ref{Raman2}, we
expect three domains in a real sample, which may suppress the
angular dependence.  It will be an experimental challenge to favor a
single-domain by applying a perturbation of the same symmetry.
\end{thebibliography}
\end{document}